# Does the Catalog of California Earthquakes, with Aftershocks Included, Contain Information about Future Large Earthquakes?

by

John B Rundle[1,2,3,6] Andrea Donnellan[4],

Geoffrey Fox[5], Lisa Grant Ludwig[6] and James Crutchfield[1,3]

[1] Department of Physics
University of California, Davis, CA

[2] Department of Earth and Planetary Science
University of California, Davis, CA

[3] Santa Fe Institute
Santa Fe, NM

[4] Jet Propulsion Laboratory
California Institute of Technology
Pasadena, CA 91109

[5] University of Virginia
Charlottesville, VA

[6] University of California
Irvine, CA

## Abstract

Yes

## Key Points

- Interval statistics have been used to conclude that major earthquakes are random events in time and cannot be anticipated or predicted

- Machine learning is a powerful new technique that enhances our ability to understand the information content of earthquake catalogs

- We show that catalogs contain significant information on current hazard and future predictability for large earthquakes

## Plain Language Summary

The question of whether earthquake occurrence is random in time, or perhaps chaotic with order hidden in the chaos, is of major importance to the determination of risk from these events. It was shown many years ago that if aftershocks are removed from the earthquake catalogs, what remains are apparently events that occur at random time intervals, and therefore not predictable in time. In the present work, we enlist machine learning methods using Receiver Operating Characteristic (ROC) analysis. With these methods, probabilities of large events and their associated information value can be computed. Here information value is defined using Shannon Information Entropy, shown by Claude Shannon (Shannon, 1948) to define the surprise value of a communication such as a string of computer bits. Random messages can be shown to have high entropy, surprise value, or uncertainty, whereas low entropy is associated with reduced uncertainty and high reliability. An earthquake nowcast probability associated with reduced uncertainty and greater reliability is most desirable. Examples of the latter could be the statements that there is a 90% probability of a major earthquake within 3 years, or a 5% chance of a major earthquake within 1 year. Despite the random intervals between major earthquakes, we find that it is possible to make low uncertainty, high reliability statements on current hazard by the use of machine learning methods.

## Introduction

Are major earthquakes random events in time? Or possibly chaotic, with order in the chaos if we know where to look? These questions lie at the heart of the debate on whether earthquakes can be predicted or anticipated, and whether it is possible to quantitatively characterize the current state of earthquake hazard.

Many years ago, Gardner and Knopoff (1974) wrote a paper with the title: "Is the sequence of earthquakes in Southern California, with aftershocks removed, Poissonian?" Their abstract: "Yes." The analysis they did was based on fitting the *intervals* between events to an exponential probability distribution, which is often called Poisson statistics. This type of statistics is well-known to apply to many types of random counting problems, from the arrivals of automobiles in parking lots, to neutron decay, to calls per hour at a call center, and many other applications.

Since that time, many other researchers have searched for temporal structure in earthquake intervals, with generally negative results (e.g., Scholz, 2019; comprehensive review by Rundle et al., 2021a and references therein). Exceptions do exist, such as are seen in Episodic Tremor and Slip and small repeating earthquakes (Rundle et al., 2021a; Rouet-Leduc, 2019), but this behavior does not generally apply to large damaging earthquakes.

In all of these studies, the fundamental question underlying these investigations can be phrased as: How much information does an earthquake catalog contain? This is the question that we consider in this paper.

To summarize our results:  We find that there is skill in earthquake nowcasts, as measured by the Receiver Operating Characteristic (ROC) curve, used in machine learning to evaluate signal detection.  Skill is defined as the ability to discriminate between true signals and false signals.  We quantify this in terms of Shannon Information Entropy, using as probabilities the ROC curve and its associated Precision (Positive Predictive Value).  We show that nowcasts of real data have lower entropy (higher information content) than random data.  Using a simple simulation of a nowcast *state variable* curve with random (exponential) recurrence times, we show that Poisson recurrence does not imply a lack of predictability or skill using the state variable.  The state variable time series resembles the long-hypothesized cycle of tectonic stress accumulation and release for major earthquakes. We conclude that the observation of Poisson recurrence statistics does not necessarily imply a lack of earthquake predictability.

## Data and Method

In recent research, we have developed methods that we call earthquake *nowcasting* whose goal is to estimate the current state of hazard.  A number of authors have now begun to use these methods in a variety of applications. Recent research has developed the idea of earthquake nowcasting, which uses state ("proxy")  variables to infer the current state of the earthquake cycle (Rundle et al., 2016, 2018, 2019, 2021a,b; Rundle et al., 2022; Rundle and Donnellan, 2020; Pasari and Mehta, 2018; Pasari, 2019, 2020; Pasari and Sharma, 2020;  Luginbuhl et al. 2019;  2020). An approach such as this is needed since the cycle of stress accumulation and release is not observable (Rundle et al., 2021b; Scholz, 2019).  These first approaches to nowcasting has been based on the concept of natural time (Varotsos et al., 2001; 2002; 2011, 2013; 2014; 2020a,b; Sarlis et al., 2018).

More specifically, in this work we analyze the result of applying a filter that, when applied to a timeseries of small earthquakes, reveals the cycle of large earthquake occurrence and recovery.  Details of the process of building, optimizing, and applying the filter is indicated in Figure 1, and discussed elsewhere (Rundle et al, 2022).  The Python code used to compute the filter is available on the ESSOAR site as well.  In this section, we sketch the process, details of which can be found in the cited references.

A critical component of the current approach is that the information is encoded in the earthquake clusters or bursts, a series of events closely spaced in time (Rundle et al., 2020; Rundle and Donnellan, 2020).  Bursts are a temporal clustering of highly correlated seismicity, typically in a small spatial region.

**Data.**  Referring to Figure 1, we begin with the seismicity in a regional box of size $10°$ latitude by $10°$  longitude centered on Los Angeles, CA (Figure 1a).  The timeseries of earthquakes in that region since 1970, having magnitudes M > 3.29, is shown in Figure 1b as a series of vertical lines. Also shown as a blue curve is the exponential moving average (EMA) with number of weights *N* = 36 [1]. Note that the blue curve shows an "inverted" cycle of large earthquakes that is the primary basis for the nowcast filter.

In Figure 1c, we show a time series for the mean number $\mu(t)$ of small earthquakes as a function of time. The mean is taken beginning in 1960, and is also shown since 1970. It can be seen that the mean does not indicate a steady state. Instead, there is a general increase in mean number of events up to about 1993, after which it shows a cycle similar to that in Figure 1b.

This catalog behavior may be due either to actual tectonic processes, or perhaps to changes in methods of earthquake detection and magnitude assignment in the early 1990's, when the network was fully automated and digital (Hutton et al., 2010). In fact, it is interesting that the temporal trends in Figure 1c seem to somewhat mirror the general historical change in number of seismic stations in California as shown in Figure 3 of Hutten et al. (2010).

**State Variable $\Theta(t)$.** The data in Figures 1b and 1c are then combined to form the state variable timeseries $\Theta(t)$ shown in Figure 1d. The state variable itself is the EMA average of the small earthquakes, then adjusted using the current mean number $\mu(2022)$ of small earthquakes, using a constant of proportionality $\lambda$. The *N*-value and $\lambda$-value are obtained by optimizing the ROC skill.

The adjustment corresponds to an assumption that there is a minimum number of small earthquakes that occur each month. An important component of this adjustment is the assumption that there appears to be a transition from unstable seismic slip, observable with seismometers, to stable sliding that is observable only with geodetic observational instruments such as GNSS or InSAR. Figure 1d then represents an "inverted" and adjusted and EMA averaged timeseries of the small earthquakes.

As noted above, the blue curve in Figure 1 shows a sudden increase at the time of large earthquakes, and thus the state variable $\Theta(t)$, which is the inverse, shows a sudden decrease.

**Receiver Operating Characteristic (ROC).** To calculate the the EMA *N*-value, and the contribution of the mean number $\mu(t)$ of small earthquakes, we construct the temporal Receiver Operating Characteristic (ROC) for a forward time window $T_W$ =3 years beyond a given time $t$ (Rundle et al., 2022). We note that other researchers are also using ROC methods in earthquake cluster analysis (Ben-Zion and Zaliapin, 2020; Zaliapin and Ben-Zion, 2022), similar in some ways to ideas in Rundle and Donnellan (2020) and Rundle et al. (2021a,b).

The ROC curve [2] is constructed by establishing a series of increasing thresholds $T_H$ in the state variable $\Theta(t)$ from low values to high values. We then consider all values of time, and a series of 4 clauses (statements). A review of these methods can be found in Jolliffe and Stephenson (2003).

For each time $t$: if a given $\Theta(t) \geq T_H$ and a large earthquake occurs within the next $T_W$ years, we classify that as true positive TP; if $\Theta(t) \geq T_H$ and no earthquake occurs within the next $T_W$ years, we classify as false positive FP; if $\Theta(t) < T_H$ and no earthquake occurs within the next $T_W$ years, we classify as true negative TN; if $\Theta(t) < T_H$ and an earthquake does occur within the next $T_W$ years, we classify as false negative FN. We then repeat this procedure over all values of threshold $T_H$.

Having TP, FP, FN, TN, we then define the true positive rate TPR or "hit rate" TP/(TP + FN); the false positive rate FPR or "false alarm rate" FP/(FP+TN). A plot of TPR against FPR defines the ROC curve, which is the red curve in Figure 1e. For future consideration, we also define the Precision, or positive predictive value as TP/(TP+FP), the fraction of predictions that turn out to be accurate. These and other quantities are described in [2].

**Supervised Machine Learning.** The area under the ROC curve is the *Skill*, which specifies the ability of the method to discriminate between true signals and false signals. The diagonal line in Figure 1e is the no skill line, equivalent to a random predictor. Note that the area under the no skill line is 0.5.

For a method to have skill, the ROC curve must either be above the diagonal line, or below it. For a method with skill, the area under the ROC curve can either be a maximum of 1.0, or a minimum of 0.0. For future reference, we define a skill index *SKI* in % as a function of the Relative Skill $R_S$ = |*Skill* - 0.5|:

$$SKI = -100 \, ( R_S \, Log_2 \, R_S \, + \, (1 - R_S) Log_2 (1 - R_S) \, ) \tag{1}$$

*SKI* can then range from 0% (when skill = 0.5), to 100% (when skill is either 1.0 or 0.0). In Figure 1e, the no skill area is indicated by the darker shaded area. The skill of the nowcasting method that we discuss here is indicated by the total shaded area.

As discussed in Rundle et al. (2022), we find the optimal values of *N* for the EMA, and the contribution of the mean earthquake $\mu(t)$ adjustment $\lambda$, by maximizing the skill. This is indicated in Figure 1d and 1e as a feedback between the state variable curve and the ROC skill calculation. This procedure results in a filter that has been optimized by well-understood, reliable methods. We note that the code is available on the AGU preprint archive ESSOAR (Rundle et al., 2022, supplemental tab).

**Shannon Information.** Claude Shannon's famous paper on statistical communication theory (Shannon, 1948) describes a measure of the information content of a message between communicating parties. It is based on the idea of viewing a message consisting of a bit string as a series of intermixed 1's and 0's, with an associated entropy of mixing. Examples of the use of these methods can be found, e.g., in Cover and Thomas (1991), and Stone (2015).

The usual interpretation of Shannon information entropy is then the number of binary yes/no questions that must be asked in order to determine the information in the message being sent. If more questions are required, the entropy is higher and the information communicated is more surprising. Conversely if fewer questions are required, the entropy is lower and the information communicated is not as surprising.

For a message in which symbols in an alphabet (i.e., the 1's and 0's) have probability mass function $p(\omega)$, where $\omega \in [0,1]$, the self-information $I_{self}$ associated with a given symbol is:

$$I_{self} = -Log_2 p(\omega) \quad (2)$$

The Shannon information of a string of symbols is then given by the expectation of the self-information:

$$I_S = -\sum_{\omega} p(\omega) Log_2 p(\omega) \quad (3)$$

Comparing (3) with the skill index *SKI*, we see that equation (1) is an information entropy-based definition.

## Results

In this section we compute the information content/entropy using the statistics of the ROC curve, and the time series precision. In Figure 2, we first address the skill and information content of the method outlined in Figure 1 for a continuum of future time windows $T_W \in [0.125, 8.5]$ years.

**Skill.** Figure 2a shows the same ROC diagram as in Figure 1e for a future time window of $T_W = 1$ year. As discussed previously, the red curve is the true positive rate (TPR), which ranges from 0 to 1. The diagonal line is the true positive rate for an ensemble of 50 random time series, each of which were obtained from the state variable time series $\Theta(t)$ using a bootstrap procedure of random sampling with replacement. The ensemble of random time series is shown as the cyan curves grouped near the diagonal line.

The skill, which is the area under the ROC curve, is shown in Figure 2b as function of the future time window $T_W$, for fixed EMA *N*-value and $\lambda$-value. Figure 2c shows the skill index *SKI* defined in (1), also as a function of $T_W$. Both Figures 2b,c indicate that there is a maximum in skill at a value $T_W$ = 0.625 years, and no skill at $T_W$ = 6.875 years, where the skill curve crosses the no-skill (dashed horizontal) line.

**Shannon Information from ROC.** To calculate the Shannon Information entropy as a function of $T_W$ using (3), we need a probability mass function *pmf*. For this purpose, we use the ROC curve as a cumulative distribution function, and difference it with respect to threshold values $T_H$ to obtain the *pmf*. Because the ROC curve was constructed using 200 values of $T_H$, there are 199 values of the *pmf* => $p(\omega)$ to be used in equation (3).

To compare the results with those for the no skill diagonal line, we note that the diagonal line can also be regarded as a cumulative distribution, but for a uniform *pmf* whose value is the constant *pmf* => $p(\omega)$ = 1/$N$. For this value of *pmf*, it is easy to show that $I_S$ = 7.64 bits.

According to the conventional interpretation of Shannon information, one would need to ask, on average, 7.64 yes/no questions to establish the value of a random state variable just prior to the occurrence of a major earthquake during the following $T_W$ years. Or in other words, the number of yes/no questions needed to determine whether a given random threshold state is followed by a window $T_W$ that contains a large earthquake.

By contrast, the actual ROC curve has a lower value of $I_S$, and therefore more information content, and lower entropy, than the random ROC (diagonal line). For the value of $T_W$ = 1.0 year, we find $I_S$ = 4.29 bits, corresponding on average to 4.29 yes/no questions.

A selection of these data are also summarized in Table 1, and are compared to data from a simple illustrative simulation discussed below. Data for skill, skill index, ROC Information, Information from random ROC, Kullback-Leibler Divergence [3], and Jensen-Shannon Divergence[4] are shown in the table as well. These latter quantities are measures of the difference in information entropy between the data and a random nowcast.

**Shannon Information from Precision (PPV).** More insight into the information content/entropy of the state variable $\Theta(t)$ can be realized using the positive predictive value (PPV) probability, or precision. Figure 3a shows the optimized state variable as a function of time, an enlarged version of Figure 1d.

Note in particular that the top area of the state variable curve corresponds to enhanced quiescence prior to the occurrence of a large earthquake, as explained previously and in Rundle et al. (2022). Conversely, the bottom area of the curve corresponds to enhanced activation, for example aftershock occurrence following a large event.

Figure 3b shows the precision, and Figure 3c shows the corresponding self information $I_{self}$, equation (2), both quantities on the horizontal axis and shown as a function of the threshold value $T_H$ on the vertical axis. These are the magenta curves in those figures. Figure 3 allows one to read horizontally and associate a value of PPV and self-information $I_{self}$ with a given value of $\Theta(t)$.

Also shown in Figures 3b,c are the PPV and $I_{self}$ for an ensemble of 50 random time series, these are the cyan curves. The mean of the cyan curves is shown as a solid black line, and the 1 $\sigma$ confidence limits are shown as dashed lines. Each random time series in the ensemble is again computed by sampling with replacement the time series $\Theta(t)$, then for each curve calculating the PPV and $I_{self}$ for that curve.

A main finding from Figure 3 is that the statistics of future time windows $T_W$ for the ensemble of random time series do not depend on the value of the threshold $T_H$. The random (uniform) probability of a future window $T_W$ containing a large earthquake is about 10%, for example. By contrast, the probability of a future time window containing a large earthquake increases dramatically as the time series $\Theta(t)$ increases from bottom of the chart (activation phase) to the top (quiescence phase).

We also see in Figure 3c that the information entropy is basically the same for the ensemble or random curves as for $\Theta(t)$ in the activation condition. Conversely, as quiescence becomes more dominant and the time of a large earthquake approaches, entropy for $\Theta(t)$ decreases and information content correspondingly increases.

We can also understand why the self-information $I_{self}$ for the random time series is approximately 3.35 bits. In the figure, we considered a series of $T_W$ = 1 year windows from 1970 to early 2022. There are thus a little more than 51 non-overlapping, independent time windows.

During this time period, there are 5 major earthquakes having magnitudes $M \geq 6.75$: M6.9 Loma Prieta; M7.3 Landers; M7.1 Hector Mine; M7.2 El Mayor Cucupah; and M7.1 Ridgecrest earthquakes. If the earthquakes were distributed randomly in time, there would be a probability of $p(\omega) = 5/51 = 0.098$ of finding a large earthquake in any of these time windows.

Thus we calculate a self-information entropy for the mean of the random ensemble curves of $I_{self} = -Log_2(5/51) = 3.35$ bits. Therefore it would take on average 3.35 yes/no questions to determine if one of these future time windows $T_W$ contains a large earthquake. Conversely, it is apparent that the self-information entropy of the PPV of $\Theta(t)$ approaches 0 as the seismically quiescence phase becomes fully developed.

The primary conclusion from these calculations is that the information content is higher in the quiescence phase of seismicity than the activation phase. Or alternatively, that the activation phase has higher entropy than the quiescence phase.

## A Simple Example

We now return to the question posed at the beginning of this paper of whether large earthquakes, which have been repeatedly found to have interval statistics that are exponentially (Poisson) distributed, nonetheless have information content about past and future events.

To show that this is not a contradiction, we consider the following simple model simulation:

- The basic state variable curve $\Theta_{sim}(t)$ is specified as the logistic function:

$$\Theta_{sim}(t) = \frac{1}{(1+exp(-t'))} \quad (4)$$

where: $t' = \frac{\Delta t}{\tau} + 6$ and $\Delta t$ is the time since the last large "earthquake."

- Failure (a "large earthquake") occurs when $\Theta_{sim}(t)$ = 0.995. At failure, we then set $\Delta t = 0$, and declare that a large "earthquake" has occurred.

- After a "large earthquake" occurs, the next value of $\tau$ is chosen from an exponential distribution whose mean is taken to be 25 "months".

- The future time window $T_W$ = 40 "months" is used to evaluate nowcast skill.

- We then progressively increase $\Delta t$ by 1 "month" intervals until the next large "earthquake" occurs, at which point we repeat the process.

The results of a long simulation of 183 large "earthquakes" is shown in Figure 4. There we see that a short segment of the time series $\Theta_{sim}(t)$ as shown in Figure 4a is generally similar to that shown in Figure 3a. Figure 3b shows that there is significant nowcast skill, equal to 0.88, with a skill index of 95.81%, meaning that the true signals can be differentiated from false signals with a high degree of reliability.

By construction, however, we also see from Figures 4c and 4d that the interval statistics for $\Theta_{sim}(t)$ conform to those of a Poisson interval distribution (exponential distribution).

We can therefore say that for this model simulation, the existence of Poisson interval statistics *does not* imply a lack of predictability for $\Theta_{sim}(t)$, similar to what we found with the California data set.

## Conclusion

In this paper we have analyzed earthquake catalogs to understand the information they contain. Interval statistics observed in catalogs are usually taken to indicate random (Poisson) events having no memory. However, we have shown that the *temporal clustering,* or variation in monthly rate of the small earthquakes, does contain important information.

This temporal clustering is in the form of bursts of activity that can be modeled with invasion percolation networks (Rundle et al., 2020). We thus find that the process of de-clustering a catalog is shown to remove information content, and to increase the information entropy.

The current results are consistent with the cluster analysis of Rundle and Donnellan (2020). We also note that this general decrease in monthly rate leading up to the next big earthquake might also be regarded as the "long tail" of the Omori aftershock distribution.

We have used this idea to construct a state variable $\Theta(t)$ by defining a 2-parameter filter, based on an exponential moving average (EMA) of small earthquake seismicity, together with an assumption about the minimum number of small earthquakes during a month-long interval.

The interplay between seismic activation, for example aftershocks, and seismic quiescence, can be analyzed by standard methods. These methods are receiver operating characteristic (ROC), positive predictive value (PPV), and Shannon information entropy. We note that quiescence has been identified as a precursor to major earthquakes in previous research (Kanamori, 1981; Wyss and Haberman, 1988; Haberman, 1988; Main and Meredith, 1991; Huang et al., 2001; Chouliaras, 2009; Weimer and Wyss, 1994; Torman et al., 2010; Rundle et al., 2011; Katsumata, 2011; Nanjo, 2020).

ROC analysis clearly shows that use of the optimized state variable $\Theta(t)$ to describe the earthquake cycle in California has nowcast skill. Skill is the ability to distinguish between future time windows $T_W$ that are likely to contain a large earthquake (true signal) and those that are not (false signal). The positive predictive value PPV can be interpreted as an indicator of the chance of a large earthquake during $T_W$.

Furthermore, the Shannon information content of both the ROC and PPV can be demonstrated to contain more information, or lower surprise value, than a random predictor. Or in other words the random predictor has higher information entropy than $\Theta(t)$.

To summarize, in reference to the original question posed by Gardner and Knopoff (1974) regarding earthquake predictability, we find the following. Their conclusion may apply to earthquake interval statistics where the ordering of temporal bursts and clustering (variation in monthly rate) has been lost through the de-clustering process, thereby increasing the information entropy in the catalog.

But if small earthquakes are used to build a state variable, to which a threshold criterion is then applied, we find that there does exist information value in the resulting state variable $\Theta(t)$. The original (non-declustered) catalog is thus found to contain significant information that can be used to compute and test earthquake probabilities without need to resort to models of stress accumulation and release, for example.

**Acknowledgements**. Research by JBR was supported under NASA grant NNX12AM22G, and by DoE grant DE- SC0017324 to the University of California, Davis. Portions of this research were also carried out at the Jet Propulsion Laboratory, California Institute of Technology under contract with NASA. Research by GCF was supported by grant NSF 2210266 CINES and is gratefully acknowledged. The authors would also like to thank Dr. Eric Kvaalen for a careful reading of the manuscript, and for suggesting clarifying changes.

**Supplementary Material.** Python code that can be used to reproduce the results of this paper can be found in the Supplemental material, or on the AGU ESSOAR preprint archive version of this paper.

**Data.** Data for this paper was downloaded from the USGS earthquake catalog for California, and are freely available there. The Python code mentioned above can be used to download these data for analysis.

**Notes**

[1] https://en.wikipedia.org/wiki/Moving_average#Exponential_moving_average (accessed 7/20/2022)

[2] https://en.wikipedia.org/wiki/Receiver_operating_characteristic (accessed 7/20/2022)

[3] https://en.wikipedia.org/wiki/Kullbac-Leibler_divergence (accessed 7/20/2022)

[4] https://en.wikipedia.org/wiki/Jensen-Shannon_divergence (accessed 7/20/2022)

**Figure 1.** a) Seismicity in a regional box of size 10° latitude by 10° longitude centered on Los Angeles, CA (Figure 1a). Large red circles represent earthquakes having magnitudes M>6.9. Smaller blue circles are earthquakes with M>5.9. b) The timeseries of earthquakes in that region since 1970, having magnitudes M > 3.29. Blue curve is the exponential moving average (EMA) with number of weights $N$ = 36 [1]. c) Time series for the mean number $\mu(t)$ of small earthquakes as a function of time. The mean is taken beginning in 1960, and is also shown since 1970. d) Optimized state variable timeseries $\Theta(t)$. State variable is the EMA average of the small earthquakes, then adjusted using the current mean number $\mu(2022)$ of small earthquakes, using a constant of proportionality $\lambda$. e) The $N$-value and $\lambda$-value are obtained by optimizing the ROC skill, which is shown as the total area under the red curve. Skill for the random time series is shown as the area under the diagonal line, thus random skill = 0.5.

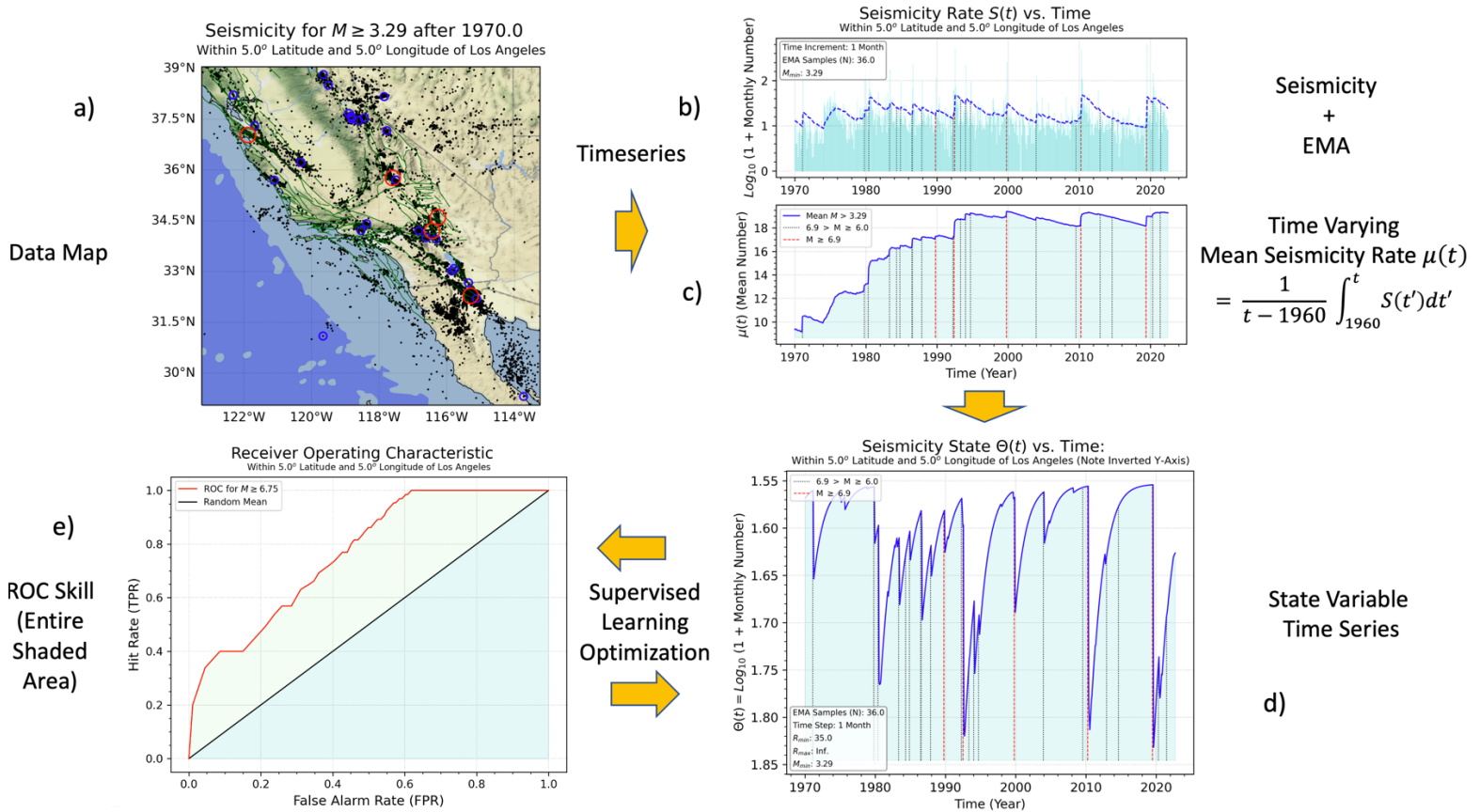

**Figure 2**. a) Shows the same ROC diagram as in Figure 1e for a future time window of $T_W = 1$ year. ROC is the red curve, representing a plot of the true positive rate (hit rate) as a function of the false positive rate (false alarm rate). The diagonal line is the true positive rate for an ensemble of 50 random time series, each of which were obtained from the state variable time series $\Theta(t)$ using a bootstrap procedure of random sampling with replacement. The ensemble of random time series is shown as the cyan curves grouped near the diagonal line. b) Shows the skill, as a function of the future time window $T_W$, for fixed EMA $N$-value and $\lambda$-value. c). Shows the skill index $SKI$ defined in equation (1), also as a function of $T_W$. d). Shows the Shannon information entropy, equation (3), as a function of future time window $T_W$. Here the information is computed from the probability mass function associated with the ROC curve. Horizontal dashed line is the information entropy for the random ROC curve (diagonal line), assuming $N = 200$ threshold values.

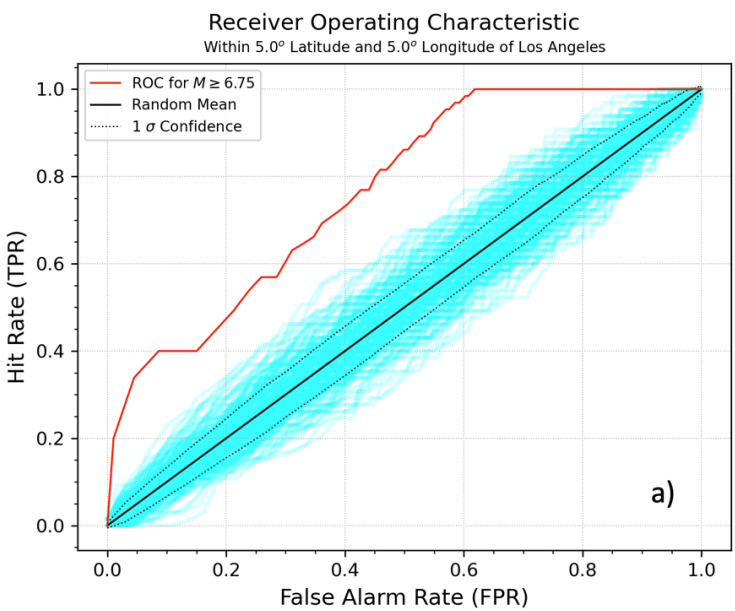
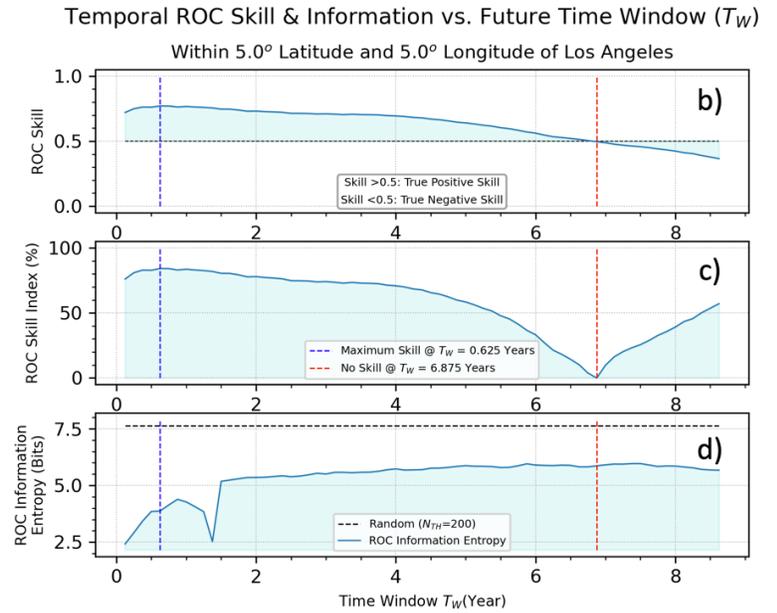

**Figure 3.** a) Shows the optimized state variable as a function of time, an enlarged version of Figure 1d. b) Shows the Positive Predictive Value, PPV or Precision. Red cuve is the PPV for the state variable shown in a), where the vertical axis is the threshold $T_H$. The cyan lines represent the PPV for 50 random time series. Mean of the time series is the solid black line, and $1\sigma$ confidence is shown as the dashed lines. c). Red curve is the corresponding self information $I_{self}$, equation (2), on the horizontal axis as a function of the threshold value $T_H$ on the vertical axis. Again, the cyan curves are the self-information for the ensemble of 50 random time series, with mean (solid black line) and $1\sigma$ confidence as the dashed lines.

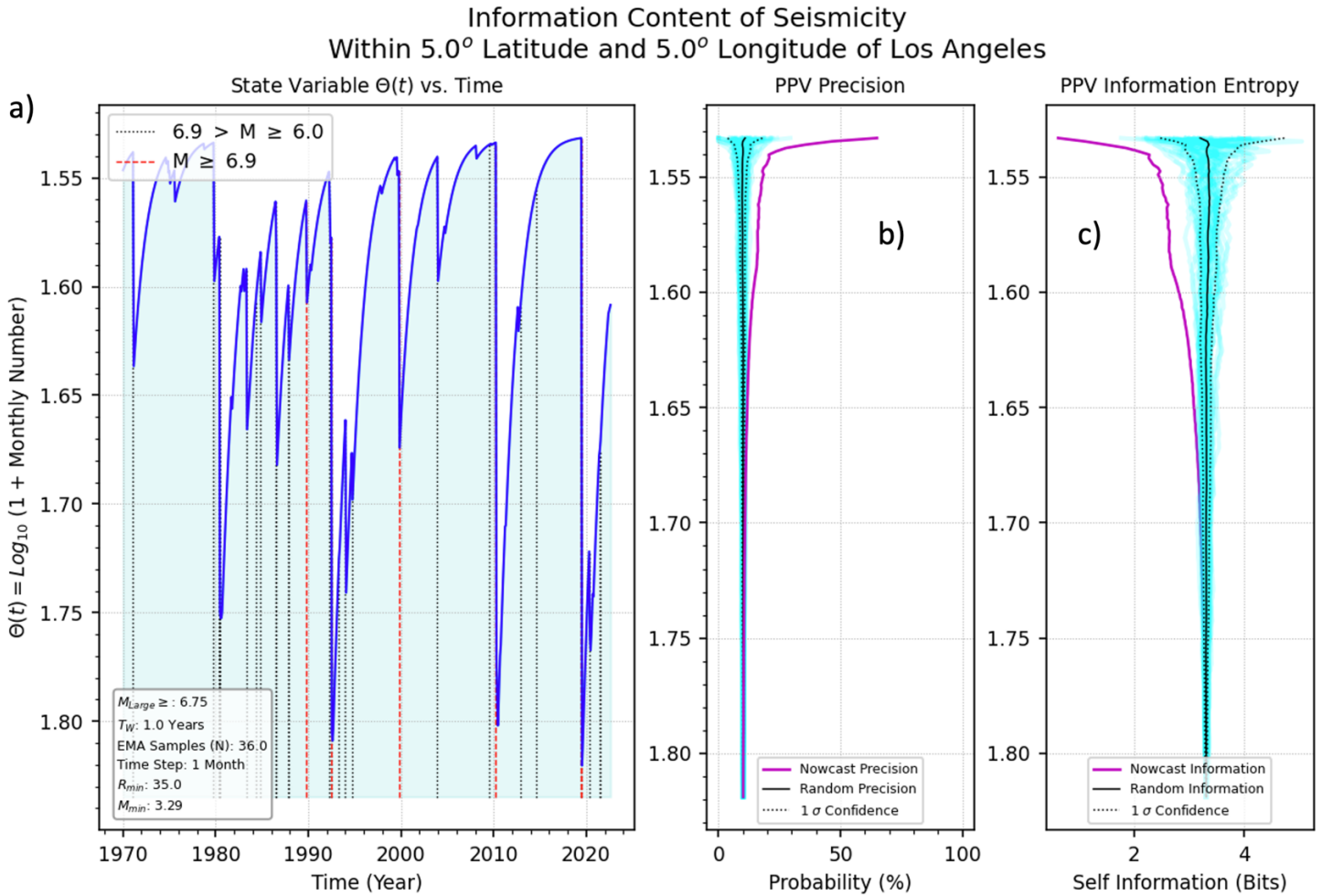

**Figure 4.** Results of a long simulation of 183 large "earthquakes". We have constructed a time series $\Theta_{sim}(t)$ using equation (4), which yields results generally similar to those in Figures 1 and 3. a) Time series $\Theta_{sim}(t)$ as a function of "time" in "months" on the left, PPV on the right. Compare to Figure 3. The vertical red line at bottom of the time series is the large "earthquake", the dashed blue line is the derivative of the time series representing the activity. On the left is the time series, on the right is the associated Precision (PPV). b) ROC curve for the time series as discussed in the text. Area of 0.88 under the ROC curve is larger than 0.5, indicating skill. Cyan curves are the skill from 50 random time series. c) Histogram of intervals between 183 large "earthquakes". d) Cumulative interval statistics, obtained from integrating histogram in c). Also shown is the dashed curve for Poisson (exponential) statistics having the same mean as the time series.

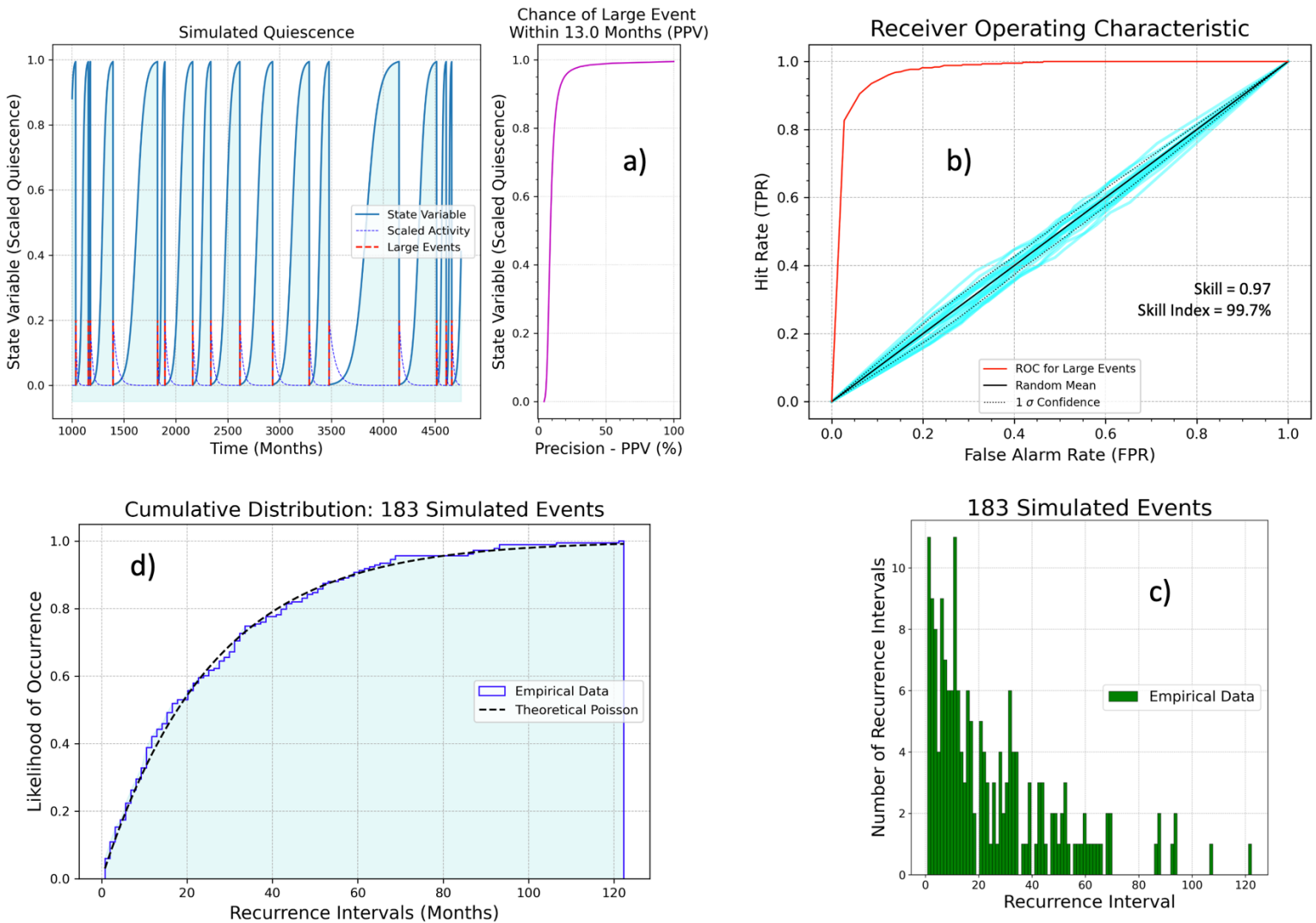

**Table 1.** Data for skill, skill index, ROC Information, Information from random ROC, Kullback-Leibler Divergence [3], and Jensen-Shannon Divergence[4]. The latter two terms refer to evaluating the distance in information space between the ROC from the filtered data and a random ROC curve (diagonal line on the ROC diagram). While both divergence quantities measure the difference in entropy between the two distributions, the Jensen-Shannon is the only one that represents a true metric. The top 4 rows of data in the table are from the California data, whereas the bottom row is from the simulation discussed in the text.

| $T_W$ | Skill | Skill Index | $I_{ROC}$ (Bits) | $I_{random}$ (Bits) | $JS_{Div}$ (Bits) | $KL_{Div}$ (Bits) |
|---|---|---|---|---|---|---|
| | | | California Data | | | |
| 1 Years | 0.77 | 83.6 % | 4.29 | 7.64 | 0.71 | 3.36 |
| 3 Years | 0.71 | 74.1 % | 5.28 | 7.64 | 0.55 | 2.36 |
| 5 Years | 0.64 | 58.7 % | 5.80 | 7.64 | 0.47 | 1.84 |
| 7 Years | 0.49 | 9.7 % | 6.34 | 7.64 | 0.35 | 1.30 |
| | | | Simulation Data | | | |
| 13 "Months" | 0.97 | 99.7 % | 1.27 | 7.64 | 0.9 | 6.37 |